# *Extraction of inherent polarization modes from a single light beam*

Xiaoyu Weng[1]*, Yu Miao[2], Qingli Zhang[2], Guanxue Wang[2], Yue Li[2], Xiumin Gao[2]* and Songlin Zhuang[2]

*1 College of Physics and Optoelectronic Engineering, Shenzhen University, Shenzhen, 518060, China.*
*2 Engineering Research Center of Optical Instrument and System, Ministry of Education, Shanghai Key Lab of Modern Optical System, School of Optical-Electrical and Computer Engineering, University of Shanghai for Science and Technology, 516 Jungong Road, Shanghai 200093, China.*
* Correspondence and requests for materials should be addressed to X.W. (email: xiaoyu@szu.edu.cn) or to X.G. (email: gxm@usst.edu.cn).

# Abstract:

Superposition of two independent orthogonally polarized beams is a conventional principle of creating a new light beam. Here, we intend to achieve the inverse process, namely, extracting inherent polarization modes from a single light beam. However, inherent polarization modes within a light beam are always entangled so that a stable polarization is maintained during propagation in free space. To overcome this limitation, we report an approach that breaks the modulation symmetry of a light beam, thereby disentangling the inherent polarization modes. Using polarization mode competition along with an optical pen, polarization modes are extracted at will in the focal region of an objective lens. This work demonstrates polarization mode extraction from a light beam, which will not only provide an entirely new principle of polarization modulation but also pave the way for multidimensional manipulation of light fields, thereby facilitating extensive developments in optics.

# Introduction

As a natural property of light beams, polarization represents the inherent oscillation of the electric field. Light beams with special polarization possess unique properties [1-5] that can not only deepen our understanding of light beams but also play a vital role in many optical applications, including optical imaging [6], optical trapping [7, 8], optical communication [9, 10] and optical lithography [11-13]. Generally, every light beam can be considered a combination of two independent orthogonally polarized beams. Throughout the development of optics, superposing two orthogonally polarized beams has been a conventional principle of creating a light beam with arbitrary polarization [14-18], which has already been written about in the optics textbooks. For example, the combination of x and y linearly polarized beams can give rise to a circularly polarized beam, while a radially polarized beam can be realized by overlapping left and right circularly polarized beams [9]. However, we wonder whether the inverse process of the above conventional principle can be realized in free space. More intuitively, can inherent polarization modes be extracted from a single light beam?

A pair of two particular orthogonal polarization modes can create a light beam with arbitrary polarization. However, this light beam contains not just the above pair of polarization modes but also many other pairs of polarization modes. Taking a radially polarized beam as an example, it can be divided into a pair of x and y linear polarization modes or a pair of left and right circular polarization modes. Adding both pairs of polarization modes together does not affect the whole polarization state of the radially polarized beam. In other words, not only a pair of x and y linear polarization modes but also other pairs of polarization modes simultaneously exist within a radially polarized beam, such as a pair of left and right polarization modes. Suppose that multiple pairs of inherent polarization modes exist within a light beam. All inherent polarization modes are intertwined with each other so that the light beam satisfying the wave function can maintain a stable polarization during propagation in free space. Thus, from the viewpoint of physics optics, the entanglement of inherent polarization modes within a light beam makes polarization mode extraction impractical. However, the solution to this physics problem would undoubtedly represent an entirely new principle of polarization modulation similar to that of overlapping two independent light beams.

Here, we report an approach that overcomes the entanglement of inherent polarization modes by breaking the modulation symmetry to extract an arbitrary polarization mode from a single light beam in the focal region of an objective lens. As a generalized property of light beams, the modulation symmetry is broken by adjusting the weight factors of inherent polarization modes, thereby disentangling the inherent polarization modes in free space. Using the polarization mode competition along with an optical pen [19], multiple polarization modes are extracted at will in the focal region of an objective lens. Polarization mode extraction from a single light beam provides an entirely new principle of polarization modulation, which may fundamentally influence optical applications and deepen our understanding of light beams.

# Results

**Modulation symmetry of a light beam**

In classical optics, a light beam satisfying the wave function always maintains an unchanged polarization when propagating in free space. The stable polarization therefore requires that the inherent polarization modes are intertwined. Otherwise, the polarization of the light beam cannot be maintained in free space. Whether the inherent polarization modes of light beams can be extracted in free space is quite questionable. The key to solving this problem requires a new understanding of the entanglement of polarization modes.

According to Supplementary Equations (2) and (3), a light beam can be composed of left and right circular polarization modes in a rectangular coordinate system or radial and azimuthal polarization modes in a cylindrical coordinate system. These orthogonal polarization modes are modulated by different factors. For example, the right and left circular polarization modes $|\mathbf{R}\rangle$ and $|\mathbf{L}\rangle$ are modulated by $\exp(\pm imf(\varphi,\theta))$, and the polarization modes $|\mathbf{RVB}_\sigma\rangle$ and $|\mathbf{LVB}_\sigma\rangle$ are modulated by the factors $\cos nf(\varphi,\theta)$ and $\sin nf(\varphi,\theta)$. However, these pairs of inherent polarization modes, regardless of the coordinate system, possess identical modulation factors $mf(\varphi,\theta)$ and $nf(\varphi,\theta)$, respectively. Here, we call this property of inherent polarization modes the modulation symmetry of a light beam. Because of the modulation symmetry, the inherent polarization modes are always entangled, thereby maintaining an identical polarization of the entire light beam during propagation in free space, as shown in Figure 1. That is, the entanglement of polarization modes is caused by the modulation symmetry of the light beam.

**Breaking the modulation symmetry**

As a generalized property of light beams, the modulation symmetry is the physical basis of the entanglement of polarization modes, which makes extracting one polarization mode from the others impossible. For this reason, the key to extracting a polarization mode from a light beam lies in how to break the modulation symmetry without affecting the inherent target polarization mode.

However, not all light beams are suitable for breaking the modulation symmetry. Modulation asymmetry between polarization modes can only occur under the condition that the modulation factors $mf(\varphi,\theta)$ and $nf(\varphi,\theta)$ are variable (see Supplementary Note 2). That is, the light beam must be inhomogeneously polarized. Otherwise, when $mf(\varphi,\theta)$ and $nf(\varphi,\theta)=const$, the light beam in Supplementary Equation (1) represents a homogeneously polarized beam. All polarization modes within the light beam are always modulated simultaneously, thereby making modulation symmetry breaking impossible.

To break the modulation symmetry of the light beam, we take a familiar vector beam, namely, an $m$-order vector vortex beam (VVB), as an example. Mathematically, an $m$-order VVB can be expressed as [9, 23]

$$\mathbf{E}_{mc} = \exp(im\varphi)|\mathbf{R}\rangle + \exp(-im\varphi)|\mathbf{L}\rangle, \tag{1}$$

$$\mathbf{E}_{ml} = \cos n\varphi|\mathbf{RVVB}_{\sigma}\rangle + \sin n\varphi|\mathbf{LVVB}_{\sigma}\rangle, \tag{2}$$

where $|\mathbf{L}\rangle = \begin{bmatrix} 1 & i \end{bmatrix}$ and $|\mathbf{R}\rangle = \begin{bmatrix} 1 & -i \end{bmatrix}$ denote the left and right circularly polarized modes, respectively, and $\varphi$ is the azimuthal angle. $|\mathbf{RVVB}_{\sigma}\rangle = [\cos\sigma\varphi \quad \sin\sigma\varphi]^{T}$ and $|\mathbf{LVVB}_{\sigma}\rangle = [-\sin\sigma\varphi \quad \cos\sigma\varphi]^{t}$ are two $\sigma$-order VVBs with orthogonal polarization. $t$ denotes the matrix transpose operator. Here, $m = n + \sigma$.

After modulation by $T = \exp i(n\varphi - \beta)$ and $T = \cos(n\varphi - \beta)$, the $m$-order VVBs in Equations (1) and (2) turn into

$$\mathbf{E}_{mc} = \exp i[(m+n)\varphi + \beta]|\mathbf{R}\rangle + \exp i[(-m+n)\varphi + \beta]|\mathbf{L}\rangle, \tag{3}$$

$$\mathbf{E}_{mv} = |\mathbf{RVVB}_{2n+\sigma,-\beta}\rangle + |\mathbf{RVVB}_{\sigma,\beta}\rangle, \tag{4}$$

respectively. Here, $|\mathbf{RVVB}_{2n+\sigma,-\beta}\rangle$ and $|\mathbf{RVVB}_{\sigma,\beta}\rangle$ are $2n+\sigma$- and $\sigma$-order VVBs with polarization directions of $-\beta$ and $\beta$, respectively, which can be written as

$$|\mathbf{RVVB}_{2n+\sigma,-\beta}\rangle = [\cos((2n+\sigma)\varphi - \beta) \quad \sin((2n+\sigma)\varphi - \beta)]^{T}, \tag{5}$$

$$|\mathbf{RVVB}_{\sigma,\beta}\rangle = [\cos(\sigma\varphi + \beta) \quad \sin(\sigma\varphi + \beta)]^{T}. \tag{6}$$

Equations (1) and (2) represent two classical forms of an $m$-order VVB: one is a pair of circular polarization modes, and the other is a pair of linear polarization modes. For the latter one in Equation (2), the $m$-order VVB can be realized by two orthogonally polarized VVBs with identical order $\sigma$, namely, $|\mathbf{LVVB}_{\sigma}\rangle$ and $|\mathbf{RVVB}_{\sigma}\rangle$. In the case of $\sigma = 0$, $|\mathbf{LVVB}_{\sigma}\rangle$ and $|\mathbf{RVVB}_{\sigma}\rangle$ are simplified into x and y linear polarization modes, respectively. That is, the entire $m$-order VVB turns into the superposition of x and y linear polarization modes with modulations of $\cos m\varphi$ and $\sin m\varphi$, respectively. Because of the identical modulation orders $m$ and $n$ in Equations (1) and (2), both pairs of polarization modes are intertwined, thereby leading to the stable polarization of the $m$-order VVB in free space. However, when the $m$-order VVBs are modulated by $T = \exp i(n\varphi - \beta)$ and $T = \cos(n\varphi - \beta)$, the modulation symmetries in Equations (1) and (2) are broken. Specifically, two circular polarization modes with different topological charges $m+n$ and $-m+n$ are obtained in Equation (3), while the $m$-order VVB in Equation (4) is split into $2n + \sigma$ - and $\sigma$ -order VVBs. The breaking of the modulation symmetry brings the prospect of disentangling the inherent polarization modes of light beams, thereby providing the possibility of extracting polarization modes from a single light beam.

**Critical conditions for polarization mode extraction**

Actually, the modulation symmetry can be simply broken by a radially or azimuthally polarized beam modulated by many phases, such as the vortex phase [2, 20] or π-phase [21]. Nevertheless, extracting one polarization mode from a light beam cannot yet be realized. The reason is that the inherent polarization modes within a radially or azimuthally polarized beam overlap with each other even in the focal region of an objective lens. To extract polarization modes from a light beam, two critical conditions still need to be satisfied. One is that the polarization modes are spatially separated in free space; the other is that only the target polarization mode is retained, while its counterpart can be neglected.

For the first condition, once the modulation symmetries are broken (see Equations (3) and (4)), the $m$-order VVB cannot maintain its original polarization state any longer. Theoretically, both orthogonal polarization modes not only come in pairs but also propagate coaxially in free space. As the light beam propagates in free space, the polarization modes with larger modulation factors or higher orders, namely, $\exp i[(m+n)\varphi + \beta]|\mathbf{R}\rangle$, $\exp i[(-m+n)\varphi + \beta]|\mathbf{L}\rangle$ and $\left|\mathbf{RVVB}_{2n+\sigma,-\beta}\right\rangle$, are located at the outer ring, while the polarization modes $|\mathbf{L}\rangle$, $|\mathbf{R}\rangle$ and $\left|\mathbf{RVVB}_{\sigma}\right\rangle$ are located at the inner ring (see Supplementary Figure 3). The longer the propagation distance of the light beam is, the larger the spatial separation of polarization modes. Generally, one can simply obtain the largest spatial separation by focusing the $m$-order VVB with an objective lens.

For the second condition, the $m$-order VVB divides into two polarization modes when modulated by $T = \exp i(n\varphi - \beta)$ and $T = \cos(n\varphi + \beta)$. Thus, the two polarization modes are of equal light intensity. Although we cannot adjust the total light intensity of polarization modes, their corresponding energy densities can be tuned at will by the orders $m$ and $n$ and by $\delta$. Take the polarization modes $|\mathbf{L}\rangle$ and $|\mathbf{R}\rangle$ in Equation (3) as an example. Suppose $m, n > 0$ and $|\mathbf{R}\rangle$ is modulated by a vortex phase with a topological charge of $m+n$, while the topological charge of the vortex phase modulating $|\mathbf{L}\rangle$ is $-m+n$. Because of this topological charge difference, the divergence degrees of $|\mathbf{L}\rangle$ and $|\mathbf{R}\rangle$ differ. Generally, the larger the absolute value of the topological charge is, the higher the divergence degree. A higher divergence degree gives rise to a lower energy density of the polarization mode. For this reason, the energy density difference between the two polarization modes $|\mathbf{L}\rangle$ and $|\mathbf{R}\rangle$ can be enlarged with the increment of $\eta = \big\| m+n \big| - \big| -m+n \big\|$. Once $\eta$ increases to a certain value, the energy density of the target polarization mode $|\mathbf{L}\rangle$ or $|\mathbf{R}\rangle$ is much larger than that of its counterpart. In this way, one can obtain one polarization mode by simply making the other mode negligible. Here, we call this phenomenon polarization mode competition within the VVB. Likewise, $|\mathbf{LVVB}_\sigma\rangle$ and $|\mathbf{RVVB}_\sigma\rangle$ can also be extracted from the $m$-order VVB with an appropriate $\eta$.

**Polarization mode extraction from a VVB**

Taking the above two critical conditions into consideration, we verify polarization mode extraction using an $m=30$-order VVB in a focusing system. To achieve multiple polarization modes, the modulation factors $T = \exp i(n\varphi - \beta)$ and $T = \cos(n\varphi + \beta)$ are combined with the optical pen developed in our previous work in Ref [19]. Therefore, the overall phase of the $m=30$-order VVB can be expressed as [19]

$$\psi = \text{Phase}\left\{ \sum_{j=1}^{N} \Big[ T_j \bullet \text{PF}(s_j, x_j, y_j, z_j, \delta_j) \Big] \right\}, \tag{7}$$

where $N$ indicates the number of foci; $x_j$, $y_j$, and $z_j$ denote the position of the $j$-th focus in the focal region; and $s_j$ and $\delta_j$ are the parameters that can be used to adjust the amplitude and phase of the $j$-th focus, respectively. $T_j$ represents the modulation factor of polarization modes, where $T = \exp i(n\varphi - \beta)$ for $|\mathbf{L}\rangle$ and $|\mathbf{R}\rangle$ in the rectangular cylindrical coordinate system and $T = \cos(n\varphi + \beta)$ for $|\mathbf{RVVB}_{\sigma,\beta}\rangle$ in the cylindrical coordinate system. Note that the amplitude factor of $T = \cos(n\varphi + \beta)$ is neglected because of its small impact on the polarization modes (see Supplementary Note 3).

**Experiment**

Following the general focusing theory in Supplementary Equation (12), the inherent polarization modes of an $m$-order VVB can be extracted in the focal region by modulating the light beam with $T = \exp i\psi$. Notably, the modulation symmetry is broken in free space once the $m$-order VVB is modulated by $T = \exp i\psi$. If only the light beam propagates sufficiently, then one can always achieve polarization mode extraction in free space, not just in a focusing system. Here, we only take numerical aperture (NA)=0.01 as an example in this paper. Figure 2 presents the schematic of the experimental setup, which is formed by two optical systems. One is a VVB creation system, indicated by the green dotted box; the other is a focusing system that is formed by an objective lens (OL), indicated by the purple dotted box. A collimated incident x linearly polarized beam with a wavelength of 633 nm passes through a phase-only spatial light modulator (SLM) and two lenses ($L_3$, $L_4$) and is converted into an $m$=30-order VVB by a vortex polarizer (VP). The $m$=30-order VVB output from the first system is focused by the OL with NA=0.01 in the focusing system, and the light intensity of the polarization mode is recorded using a CCD. Here, the VP is realized by the Q-plate technique [22-24] and conjugated with the phase-only SLM by the 4f system with $L_3$ ($f_3$=150 mm) and $L_4$ ($f_4$=150 mm). Therefore, the phase of the $m$=30-order VVB in Equation (7) can be adjusted at will by the phase-only SLM.

Figure 3 presents the experimental result of a single polarization mode extraction from the $m$=30-order VVB in the focal region of the OL. Two kinds of polarization modes are extracted from the VVB for different coordinate systems. Polarization modes $|\mathbf{L}\rangle$ and $|\mathbf{R}\rangle$ in Figures 3 (b, g) are created by the vortex phase $T = \exp i(n\varphi - \beta)$ with $n = \pm 30$, as shown in Figures 3 (a, f), which are the eigenmodes of the Cartesian coordinate system. In the cylindrical coordinate system, polarization modes $|\mathbf{RVVB}_{\sigma,\beta}\rangle$ in Figures 3 (l, q, v) are realized by the phase $T = \exp i\psi$ with $\psi = \text{Phase}[\cos(n\varphi + \beta)]$, where $\beta = 0$ and $n = 29, 28, 20$ for Figures 3 (k, p, u), respectively. Note that the polarization state of $|\mathbf{RVVB}_{\sigma,\beta}\rangle$ can be converted into that of $|\mathbf{LVVB}_{\sigma,\beta}\rangle$ by adjusting the parameter $\beta$. All of these modes are distinguished from each other using a quarter-wave plate (indicated by the blue arrows) and a polarizer (indicated by the purple arrows), as shown in Figures 3 (c-e, h-j, m-o, r-t, w-y), which are consistent with their theoretical results in Supplementary Figure 4.

To better understand the polarization extraction from the light beam, Figure 4 presents the entire principle of polarization mode extraction from a VVB. Two key factors exist for polarization mode extraction. One is the breaking of the modulation symmetry, and the other is the polarization mode competition. Here, we take the polarization modes $|\mathbf{L}\rangle$ in Figure 3 (b) as an example to explain the breaking of the modulation symmetry and the polarization mode competition within the VVB. When modulated by the vortex phase $T = \exp i(n\varphi - \beta)$ with $n = m$=30 in Figure 3 (a), the modulation symmetry of the VVB is broken, thereby dividing the light beam into left and right polarization modes simultaneously. Although the two polarization modes are two equal parts of the $m$=30-order VVB, their energy densities are very different because the two modes possess different topological charges. Generally, a larger energy density of one polarization mode leads to a smaller energy density of the other.

Due to the above polarization mode competition, the polarization mode with a higher topological charge has a lower energy density, thereby leading to a smaller maximum light intensity at the outer ring. In contrast, the polarization mode at the inner ring possesses a lower topological charge, thereby leading to a larger maximum light intensity (see Supplementary Figure 3). For this reason, one can adjust the maximum light intensities of both polarization modes by controlling the difference in topological charge, namely, $\eta = \left\| m+n \right| - \left| -m+n \right\|$. Here, we define a parameter $\gamma = I_{m,outer} / I_{m,inner}$ to characterize the difference between the maximum light intensities of the inner ring $I_{m,inner}$ and outer ring $I_{m,outer}$. In the case of Figure 3 (b), $\eta$=60 and $\gamma = 0.0018$. When normalized, the right circular polarization mode at the outer ring can be neglected, and only the left circular polarization mode is retained in the focal region. Likewise, one can obtain arbitrary polarization modes $\left| \mathbf{RVVB}_{\sigma,\beta} \right\rangle$ with $\sigma = 1,2,10$ and $\beta$=0, as shown in Figures 3 (l, q, v), based on the phases in Figures 3 (k, p, u).

Figure 3 not only demonstrates the polarization mode extraction from an $m$-order VVB but also, more importantly, verifies that polarization can be controlled merely by the phase of the light beam. Thus, with the aid of an optical pen, one can simply extend the single polarization mode extraction in Figure 3 into multiple polarization mode extraction, as shown in Figure 5. After all, not just one pair of polarization modes but many pairs exist within a light beam. Figures 5 (a, e, i) present the experimental results of extracting 3×3 polarization mode arrays in the focal plane, showing (a) polarization modes $\left| \mathbf{RVVB}_{\sigma,\beta} \right\rangle$ with positive order $\sigma = 0,1...,7$ and a right circular polarization (RCP) mode at the position of the geometric focus; (e) polarization modes $\left| \mathbf{RVVB}_{\sigma,\beta} \right\rangle$ with negative order $\sigma = 0,-1...,-7$ and an RCP mode at the position of the geometric focus; and (i) polarization modes $\left| \mathbf{RVVB}_{\sigma,\beta} \right\rangle$ with positive and negative orders $\sigma = 0,\pm1...,\pm4$. Their corresponding phases are shown in Figures 5 (m, n, o), the parameters of which can be found in Supplementary Note 4. As shown in Figures 5 (b-d, f-h, j-l), $2\left| \sigma \right|$ petals are rotated along with the polarizer indicated by the purple arrows, which coincides with the corresponding theoretical results in Supplementary Figure 5. In addition, we also verify the function of parameter $\beta$ in the polarization mode transformation in Supplementary Figures 6 and 7.

## Discussion

In theory, a light beam with arbitrary polarization can be obtained by superposing two orthogonally polarized light beams. For example, the circularly polarized beam output from a quarter-wave plate is the combination of $o$ light and $e$ light. Similarly, a metasurface with effective birefringence can generate a light beam with a special polarization distribution based on two orthogonal electric field components $\mathbf{E}_x$ and $\mathbf{E}_y$, similar to a waveplate [17]. That is, this conventional principle always requires two orthogonal polarization components [14-18]. However, polarization mode extraction exceeds this fundamental limitation in that it does not rely on any polarization components but, conversely, extracts inherent polarization modes directly from a single light beam. For this reason, polarization mode extraction from a single light beam represents an entirely new principle of polarization modulation in classical optics.

Although two orthogonally polarized beams, such as polarization modes A and B in Figure 6, can form a new light beam, the light beam contains not just both of the above polarization modes but also many other polarization modes. Therefore, the inverse process of overlapping two light beams, namely, polarization mode extraction, can provide many more possibilities in terms of polarization modulation (see Figure 6). Specifically, different coordinate systems possess different inherent polarization modes within a light beam. We have already extracted the eigenmodes $\left|\mathbf{L}\right\rangle$, $\left|\mathbf{R}\right\rangle$ and $\left|\mathbf{RVVB}_{\sigma,\beta}\right\rangle$ for Cartesian and cylindrical coordinate systems in the focal region. One can predict that other eigenmodes can also be obtained in the focal region for other coordinate systems. In addition, polarization mode extraction has been demonstrated as the first attempt to manipulate the polarization of a light beam with its own phase. Therefore, multiple polarization modes can be extracted from a single light beam simultaneously, which is another advantage of polarization mode extraction over the above principle.

In the following, we discuss a technical factor of polarization mode extraction. Polarization mode extraction involves one important conceptual change: the modulation symmetry is the physical basis of polarization mode entanglement within a light beam satisfying the wave function. However, breaking the modulation symmetry does not ensure polarization mode extraction in the focal region [2, 20, 21]. For example, supposing $\eta=1$, one can also achieve modulation asymmetry of the light beam; however, the inner and outer polarization modes cannot be spatially separated, thereby making extraction of one polarization mode from the others impossible. The key parameter $\eta$ not only plays a vital role in adjusting $\gamma$ but also controls the distance between the outer and inner polarization modes. Generally, $\eta$ should not be too small. A small $\eta$ leads to a short distance between the outer and inner polarization modes, which not only gives rise to a large $\gamma$ but also causes interference between the two modes. In the case of $m < |n|$, $\eta = \left\| m+n \right| - \left| -m+n \right\|$ increases from 0 to $2m$ with the increment of $|n|$, while an identical $\eta = 60$ can be obtained in the case of $m < |n|$. Owing to the identical $\eta = 60$, the polarization modes within the $m$-order VVB with an order even larger than $m$ can be extracted in free space.

In conclusion, we have theoretically and experimentally demonstrated extraction of inherent polarization modes from a single light beam in the focal region of an OL by taking an $m$-order VVB as an example. The entanglement of inherent polarization modes is overcome by breaking the modulation symmetry of the VVB along with using the polarization mode competition. Polarization modes $\left| \mathbf{L} \right\rangle$, $\left| \mathbf{R} \right\rangle$ and $\left| \mathbf{RVVB}_{\sigma,\beta} \right\rangle$ for Cartesian and cylindrical coordinate systems are therefore extracted at arbitrary positions of the focal region with the aid of an optical pen. This work not only conveys the physical idea that the generalized property of all light beams satisfying the wave function, namely, modulation symmetry, is the physical basis of the entanglement of inherent polarization modes during propagation in free space but also presents an entirely new principle for polarization modulation, which will pave the way for polarization control based on the phase modulation of light beams.

## Data Availability

All data supporting the findings of this study are available from the corresponding author on request.

**Acknowledgments**

Parts of this work were supported by the National Natural Science Foundation of China (62022059/11804232) and the National Key Research and Development Program of China (2018YFC1313803).

**Author contributions**

X. Weng conceived of the research and designed the experiments. Yu Miao and Q. Zhang performed the experiments. Guanxue Wang, Yue Li and X. Weng analyzed all of the data. X. Weng and X. Gao supervised the experiments. X. Weng wrote the paper. X. Gao and S. Zhuang offered advice regarding its development. S. Zhuang directed the entire project.

**Competing interests statement**

The authors declare that they have no competing financial or nonfinancial interests to disclose.

# Figures

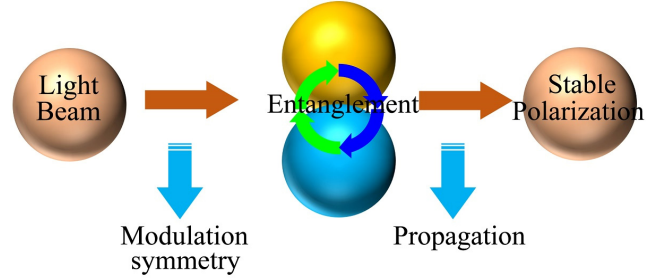

**Figure 1.** Schematic of the entanglement of inherent polarization modes. Due to the modulation symmetry, a light beam that satisfies the wave function can maintain a stable polarization during propagation in free space.

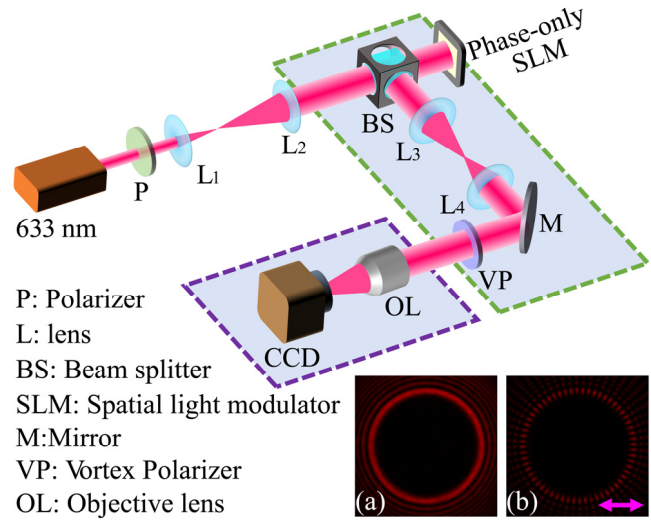

**Figure 2.** Schematic of the experimental setup for polarization mode extraction. The entire optical system is formed by the VVB creation system (the green dotted box) and a focusing system that is formed by an OL (the purple dotted box). A collimated incident $x$ linearly polarized beam with a wavelength of 633 nm can be converted into an $m$=30-order VVB by the VP, which is conjugated with the phase-only SLM by the 4f system with $L_3$ ($f_3$=150 mm) and $L_4$ ($f_4$=150 mm). The VP is a vortex polarizer realized by the Q-plate technique. The modulated VVB output from the first system is focused by the OL with NA=0.01 in the focusing system, and the light intensity of the polarization mode is recorded using a CCD. Here, subfigures (a) and (b) are the light intensities of the $m$-order VVB without and with a polarizer (purple arrows), respectively.

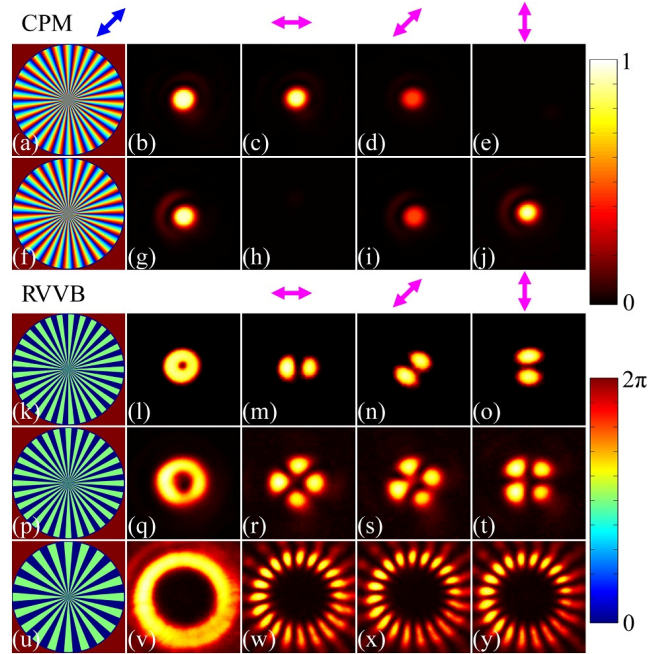

**Figure 3.** Experimental results of single polarization mode extraction. Left and right circular polarization modes (b, g) are realized based on the vortex phases in (a, f), respectively. VVBs with the order of 1, 2, and 10 (l, q, v) are extracted based on the phase $\phi = \mathrm{Phase}[\cos(n\varphi - \beta)]$, where $\beta = 0$, $n=29$ (k), $n=28$ (p), and $n=20$ (u). Subfigures (c-e, h-j, m-o, r-t, w-y) show the light intensities passing through the quarter-wave plate (indicated by the blue arrows) and the polarizer (indicated by the purple arrows). CPM and RVVB indicate the circular polarization mode and $\left|\mathbf{RVVB}_{\sigma,\beta=0}\right\rangle$, respectively. Here, all light intensities are normalized to a unit value, and the phase scale is 0-2π.

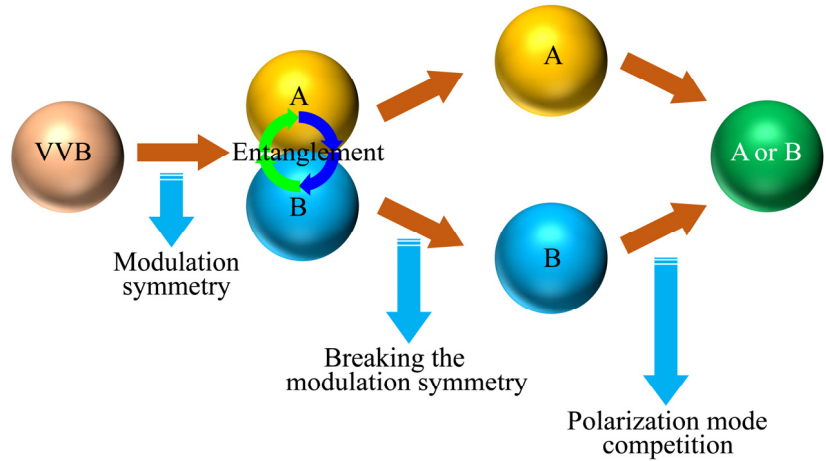

**Figure 4.** Schematic of inherent polarization mode extraction from a VVB. Here, A and B represent two arbitrary orthogonal polarization modes within the VVB. Polarization mode A or B can be extracted and even selected in the focal region of the OL by breaking the modulation symmetry as well as using polarization mode competition.

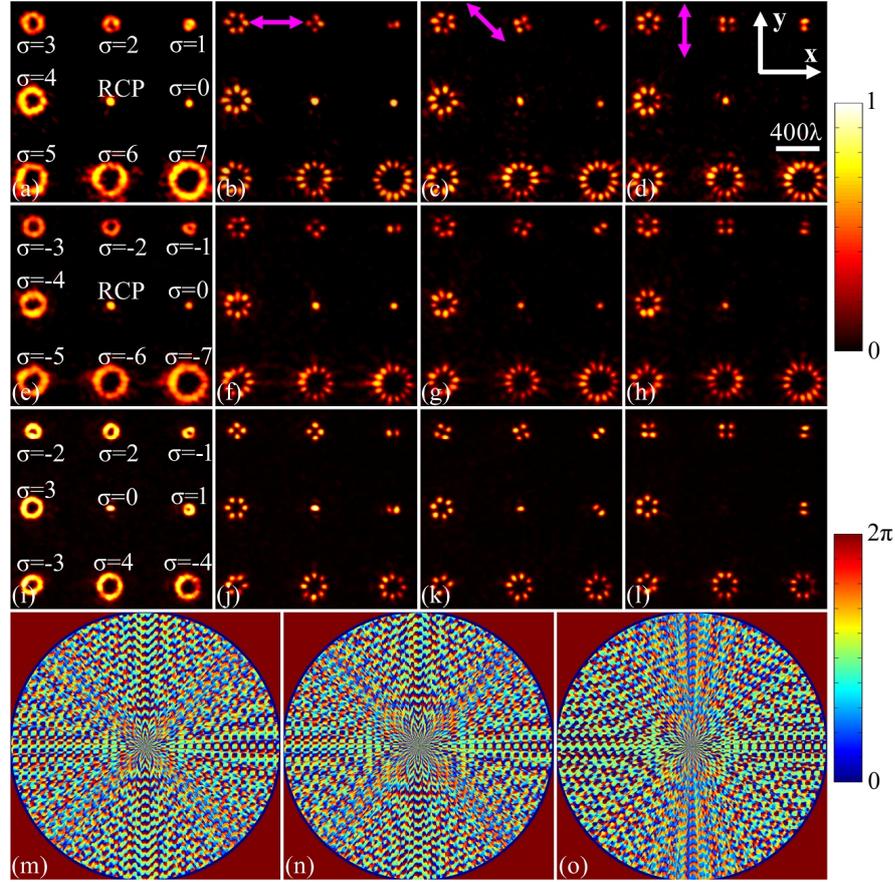

**Figure 5.** Experimental results of multiple polarization mode extraction. The 3×3 multiple polarization arrays (a, e, i) are extracted from the $m$=30-order VVB in the focal region based on the phases in (m, n, o), respectively. Here, RCP indicates a right circular polarization mode; the parameter σ is the order of polarization modes $\left|\mathbf{RVVB}_{\sigma,\beta}\right\rangle$ with $\beta$=0. Subfigures (b-d, f-h, j-l) show the light intensities of polarization modes passing through the polarizer indicated by the purple arrows. All intensities of polarization modes are normalized to a unit value, and the phase scale is 0-2π.

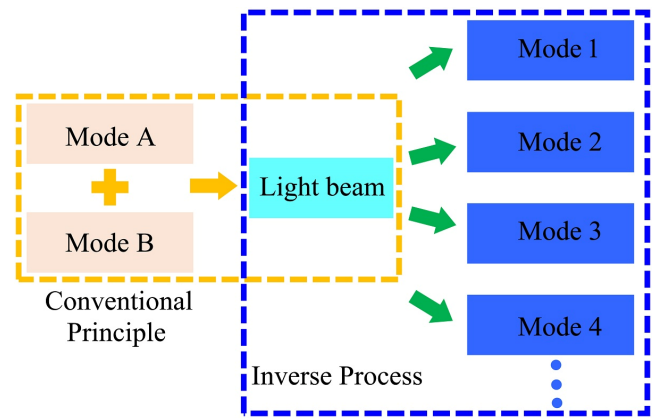

**Figure 6.** Schematic of the conventional principle and its inverse process. A light beam can be formed by overlapping polarization modes A and B. However, the inverse process of this conventional principle can provide many more possibilities. That is, many arbitrary polarization modes can be extracted from this light beam, such as polarization modes 1-4.

# Supplementary Information

## *Extraction of inherent polarization modes from a single light beam*

Weng et al.

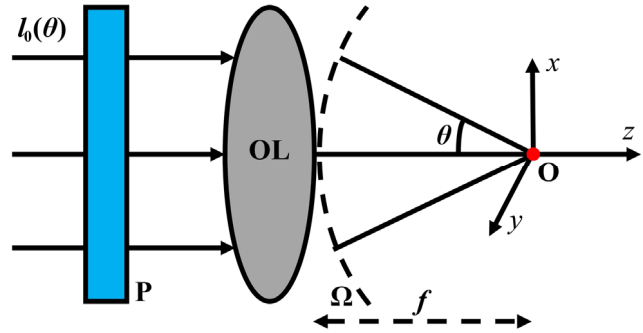

**Supplementary Figure 1** Schematic of the focusing system in Figure 2. $\Omega$ is the focal sphere, with its center at O and radius $f$, namely, the focal length of the objective lens (OL). The transmittance of the pupil filter P is expressed as Equation (7). $\theta$ is the convergent angle. $l_0(\theta)$ denotes the electric field amplitude of the incident $m$-order vector vortex beam (VVB).

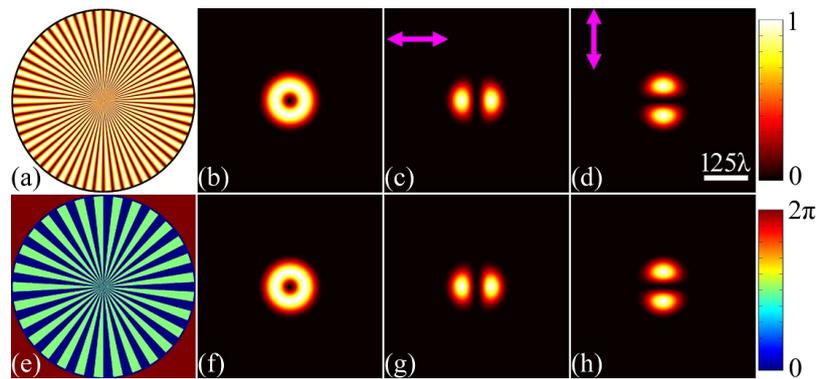

**Supplementary Figure 2.** Polarization modes generated by accurate and approximate modulation. The polarization modes $\left| \mathbf{RVVB}_{\sigma=1,\beta=0} \right\rangle$ (b and f) are extracted in the focal region of the OL by using the accurate and approximate wavefronts of the $m$=30-order VVB in (a and e), respectively. Subfigures (c, d, g, and h) show the light intensities passing through the polarizer indicated by the purple arrows. Here, the light intensities are normalized to a unit value, and the phase scale is 0-2π.

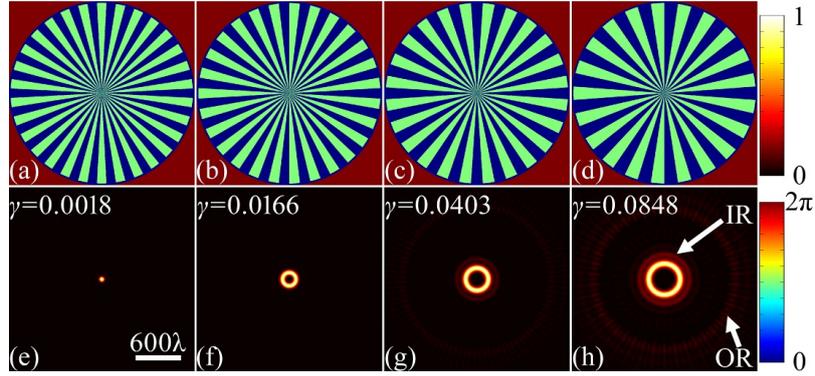

**Supplementary Figure 3.** Polarization modes with different orders. Polarization modes $\left|\mathbf{RVVB}_{\sigma,\beta=0}\right\rangle$ with the order (e) $\sigma = 0$, (f) $\sigma = 3$, (g) $\sigma = 6$, and (h) $\sigma = 9$ are extracted based on the phases in (a, b, c, and d), respectively. $\gamma = I_{m,outer} / I_{m,inner}$, where $I_{m,inner}$ and $I_{m,outer}$ are the maximum light intensities of the polarization modes in the inner ring (IR) and outer ring (OR), respectively. Here, the light intensities are normalized to a unit value, and the phase scale is 0-2$\pi$.

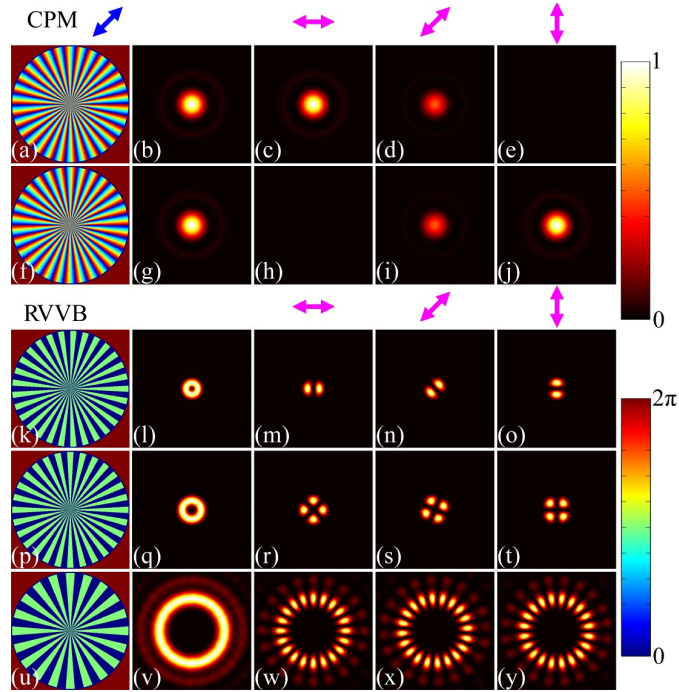

**Supplementary Figure 4.** Theoretical results of Figure 3. Left and right circular polarization modes (b, g) are realized by using the vortex phases in (a, f), respectively. VVBs with the order of 1, 2, 10 (l, q, v) are extracted by using the phase $\phi = \mathrm{Phase}[\cos(n\varphi - \beta)]$, where $\beta = 0$, $n = 29$ (k), $n = 28$ (p), and $n = 20$ (u). Subfigures (c-e, h-j, m-o, r-t, w-y) show the light intensities passing through the quarter-wave plate (indicated by the blue arrows) and the polarizer (indicated by the purple arrows). CPM and RVVB indicate the circular polarization mode and $\left|\mathbf{RVVB}_{\sigma,\beta=0}\right\rangle$, respectively. Here, all light intensities are normalized to a unit value, and the phase scale is 0-2$\pi$.

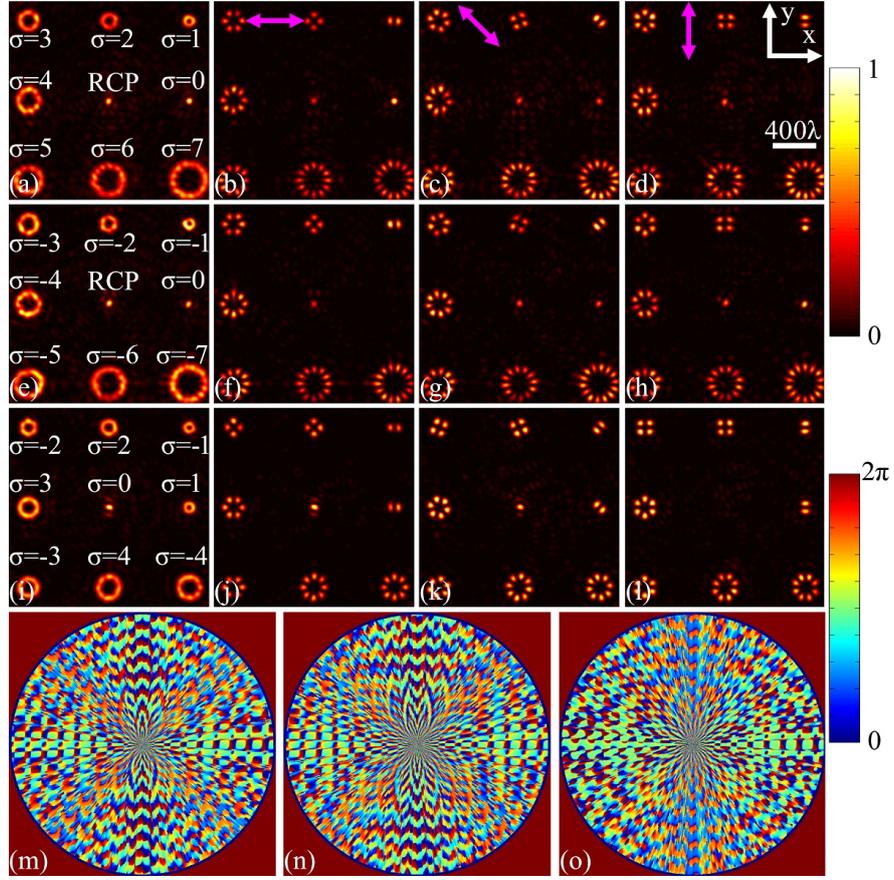

**Supplementary Figure 5.** Theoretical results of Figure 5. The 3×3 polarization mode arrays (a, e, i) are extracted from the *m*=30-order VVB in the focal region based on the phases in (m, n, o), respectively. RCP indicates a right circular polarization mode; the parameter σ is the order of polarization modes $\left| \mathbf{RVVB}_{\sigma, \beta=0} \right\rangle$. Subfigures (b-d, f-h, j-l) show the light intensities passing through the polarizer indicated by the purple arrows. The light intensities are normalized to a unit value, and the phase scale is 0-2π.

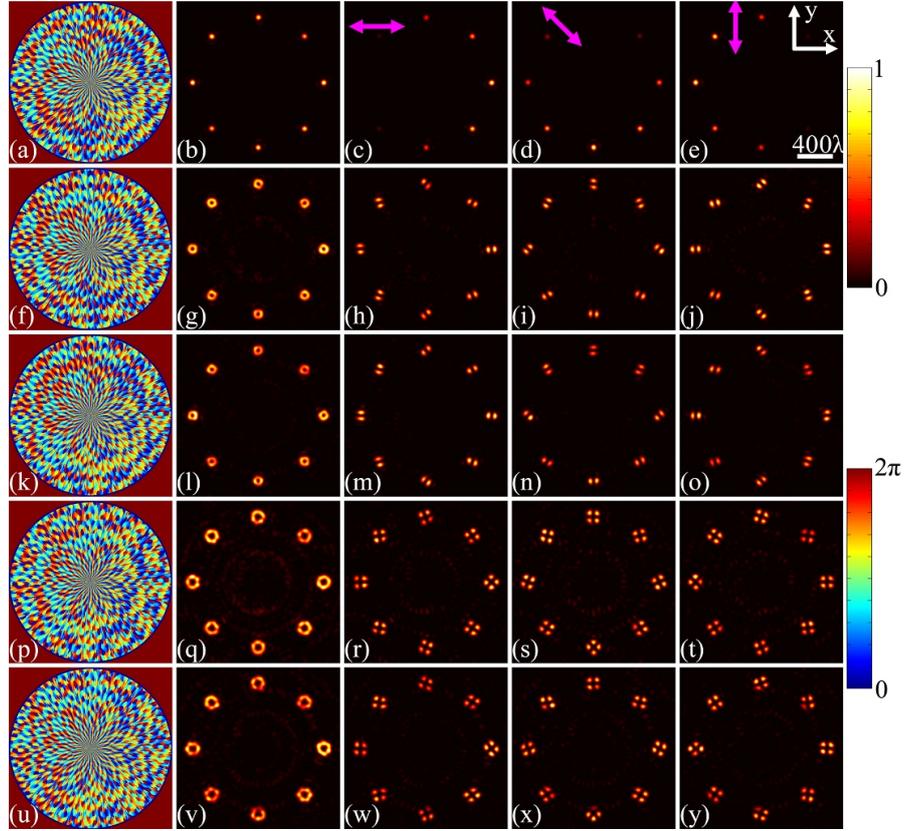

**Supplementary Figure 6.** Theoretical results of multiple polarization mode extraction with different $\beta$. Polarization modes $\left|\mathbf{RVVB}_{\sigma,\beta}\right\rangle$ with the order (b) $\sigma = 0$, (g) $\sigma = 1$, (l) $\sigma = -1$, (q) $\sigma = 2$, and (v) $\sigma = -2$ are extracted based on the phases in (a, f, k, p, and u), respectively. Subfigures (c-e, h-j, m-o, r-t, w-y) show the light intensities passing through the polarizer indicated by the purple arrows. Here, $\beta$ controls the polarization direction of each polarization mode, and all parameters of the phases in (a, f, k, p, u) can be found in Supplementary Note 4. The light intensities are normalized to a unit value, and the phase scale is 0-2$\pi$.

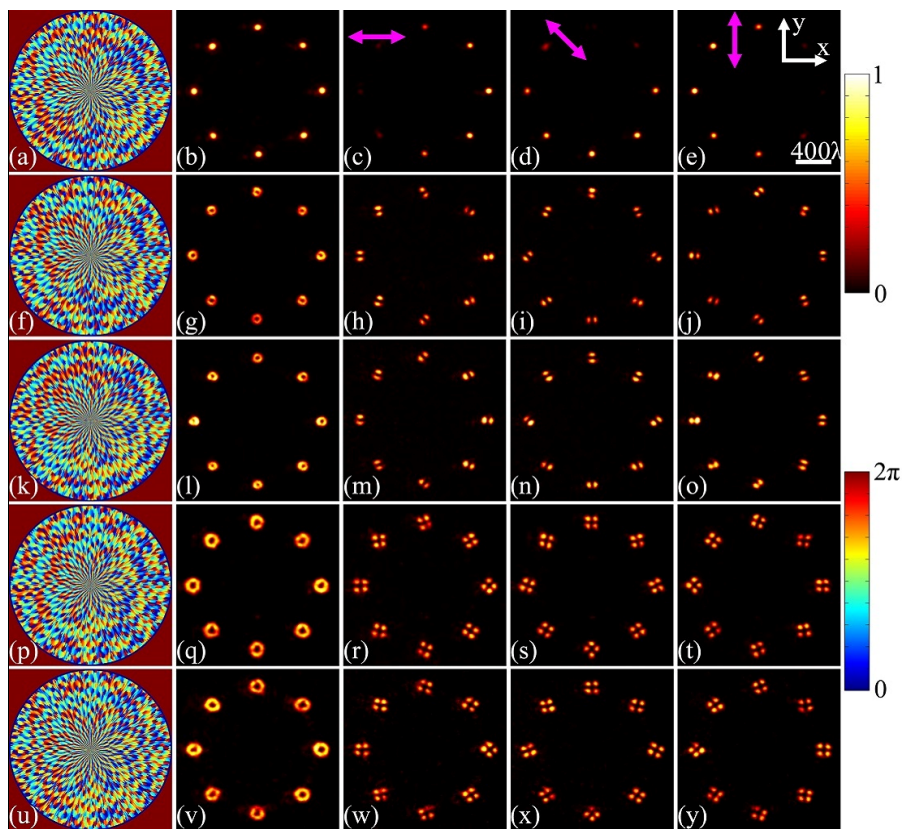

**Supplementary Figure 7.** Experimental results of Supplementary Figure 6. The light intensities of all polarization modes are normalized to a unit value, and the phase scale is 0-2π.

# Supplementary Note 1: Modulation symmetry of a light beam

In principle, a light beam can be composed of two orthogonal polarization modes, which is also the basis of a conventional interference scheme. During propagation in free space, the two polarization modes are intertwined with each other so that a stable polarization state can be maintained. Generally, a light beam with homogeneous and inhomogeneous polarization can be expressed as

$$\mathbf{E} = \begin{bmatrix} \cos mf(\varphi,\theta) & \sin mf(\varphi,\theta) \end{bmatrix}^t, \tag{1}$$

where $f(\varphi,\theta)$ represents a generalized phase function, $\theta$ and $\varphi$ are the convergence and azimuthal angles of the focusing system in Supplementary Figure 1, $t$ denotes the matrix transpose operator, and $m$ is the order of the light beam. In the case of $mf(\varphi,\theta)=\text{const}$, the light beam in Supplementary Equation (1) is a homogeneously polarized beam. For example, an x linearly polarized light beam can be obtained with $m=0$. In the case of $mf(\varphi,\theta)=\text{variate}$, the light beam possesses an inhomogeneous polarization state.

Without loss of generality, Supplementary Equation (1) can be written in two forms:

$$\mathbf{E}_{mc} = \exp\big(imf(\varphi,\theta)\big)|\mathbf{R}\rangle + \exp\big(-imf(\varphi,\theta)\big)|\mathbf{L}\rangle, \tag{2}$$

$$\mathbf{E}_{ml} = \cos nf(\varphi,\theta)|\mathbf{RVB}_\sigma\rangle + \sin nf(\varphi,\theta)|\mathbf{LVB}_\sigma\rangle, \tag{3}$$

where $|\mathbf{L}\rangle = \begin{bmatrix} 1 & i \end{bmatrix}$ and $|\mathbf{R}\rangle = \begin{bmatrix} 1 & -i \end{bmatrix}$ denote the left and right circular polarization modes, respectively, and $\varphi$ is the azimuthal angle. $|\mathbf{RVB}_\sigma\rangle = \begin{bmatrix} \cos \sigma f(\varphi,\theta) & \sin \sigma f(\varphi,\theta) \end{bmatrix}^T$ and $|\mathbf{LVB}_\sigma\rangle = [-\sin \sigma f(\varphi,\theta) \ \cos \sigma f(\varphi,\theta)]^t$ are the $\sigma$-order light beams with orthogonal polarization. $m = n + \sigma$. In the case of $\sigma = 0$, $|\mathbf{RVB}_\sigma\rangle$ and $|\mathbf{LVB}_\sigma\rangle$ are simplified into x and y linear polarization modes, respectively. That is, the $m$-order light beam in Supplementary Equation (1) can be formed by combining x and y linear polarization modes with the modulation of $\cos mf(\varphi,\theta)$ and $\sin mf(\varphi,\theta)$, respectively.

From Supplementary Equations (2) and (3), inherent polarization modes within the light beam are modulated by different weight factors. Taking Supplementary Equation (2) as an example, polarization modes $|\mathbf{R}\rangle$ and $|\mathbf{L}\rangle$ are modulated by $\exp\big(\pm imf(\varphi,\theta)\big)$. Although the signs of the two phases are inverse, their absolute values are equal, namely, $mf(\varphi,\theta)$. Likewise, polarization modes $|\mathbf{LVB}_\sigma\rangle$ and $|\mathbf{RVB}_\sigma\rangle$ modulated by $\cos nf(\varphi,\theta)$ and $\sin nf(\varphi,\theta)$ also possess an identical $nf(\varphi,\theta)$ in Supplementary Equation (3). Here, we call the above modulation of inherent polarization modes the modulation symmetry of the light beam. Due to the modulation symmetry, all inherent polarization modes within the light beam are intertwined with each other so that the polarization in Supplementary Equation (1) remains the same during propagation in free space. This is the physical basis of why one cannot extract polarization modes $\left|\mathbf{L}\right\rangle$, $\left|\mathbf{R}\right\rangle$, $\left|\mathbf{LVB}_\sigma\right\rangle$ and $\left|\mathbf{RVB}_\sigma\right\rangle$ from a single light beam in free space.

# Supplementary Note 2: Breaking the modulation symmetry of a light beam

As discussed in Supplementary Note 1, a light beam that satisfies the wave function possesses modulation symmetry between inherent polarized modes, thereby maintaining stable polarization during propagation in free space. To break the modulation symmetry, one must destroy the identical modulation factors $mf(\varphi,\theta)$ and $nf(\varphi,\theta)$. Here, we impose two particular modulation factors on the circular and linear polarization modes in Supplementary Equations (2) and (3), respectively.

For the case of Supplementary Equation (2), circular polarization modes with different modulation factors can be obtained by modulating the light beam with $\exp(\pm imf(\varphi,\theta))$, which can be expressed as

$$\mathbf{E}_{mcL} = \exp(i2mf(\varphi,\theta))|\mathbf{R}\rangle + |\mathbf{L}\rangle, \tag{4}$$

$$\mathbf{E}_{mcR} = \exp(-i2mf(\varphi,\theta))|\mathbf{L}\rangle + |\mathbf{R}\rangle. \tag{5}$$

For the case of Supplementary Equation (3), the light beam can be simplified into

$$\mathbf{E}_{mv} = \left|\mathbf{RVB}_{2n+\sigma,-\beta}\right\rangle + \left|\mathbf{RVB}_{\sigma,\beta}\right\rangle \tag{6}$$

when modulated by $\cos(nf(\varphi,\theta)-\beta)$. Here, $\left|\mathbf{RVB}_{\eta,\omega}\right\rangle = [\cos(\eta f(\varphi,\theta)+\omega) \quad \sin(\eta f(\varphi,\theta)+\omega)]^t$ is a $\eta$-order light beam with polarization direction $\omega$. Thus, $\left|\mathbf{RVB}_{2n+\sigma,-\beta}\right\rangle$ and $\left|\mathbf{RVB}_{\sigma,\beta}\right\rangle$ can be obtained with $\eta=2n+\sigma$, $\omega=-\beta$ and with $\eta=\sigma$, $\omega=\beta$, respectively. For $\beta=0,\pi/2$, the light beam in Supplementary Equation (6) can be simplified into

$$\mathbf{E}_{mvR} = \left|\mathbf{RVB}_{2n+\sigma}\right\rangle + \left|\mathbf{RVB}_{\sigma}\right\rangle, \tag{7}$$

$$\mathbf{E}_{mvL} = -\left|\mathbf{LVB}_{2n+\sigma}\right\rangle + \left|\mathbf{LVB}_{\sigma}\right\rangle. \tag{8}$$

Although polarization modes $|\mathbf{L}\rangle$, $|\mathbf{R}\rangle$, $\left|\mathbf{RVB}_{2n+\sigma,-\beta}\right\rangle$ and $\left|\mathbf{RVB}_{\sigma,\beta}\right\rangle$ are of different modulation factors or different orders in Supplementary Equations (4), (5) and (6), we still cannot assert that the modulation symmetry of the light beam is broken. For example, if $mf(\varphi,\theta)=const$ or $nf(\varphi,\theta)=const$, then the light beam in Supplementary Equation (1) is a homogeneously polarized beam, the polarization modes of which always have an identical modulation weight factor in Supplementary Equations (4), (5) and (6). For this reason, inherent polarization modes are still intertwined, and the modulation symmetry of a homogeneously polarized beam cannot be broken.

When $mf(\varphi,\theta)$ and $nf(\varphi,\theta)$ are variable, Supplementary Equation (1) becomes an inhomogeneously polarized beam. Take a radially polarized beam as example. In the case of $mf(\varphi,\theta)=\varphi$, Supplementary Equation (1) represents a radially polarized beam. When modulated by $\exp(\pm i\varphi)$ and $\cos(\varphi-\beta)$, the light beam can be simplified into

$$\mathbf{E}_{mcL}=\exp(i2\varphi)|\mathbf{R}\rangle+|\mathbf{L}\rangle, \tag{9}$$

$$\mathbf{E}_{mcR}=\exp(-i2\varphi)|\mathbf{L}\rangle+|\mathbf{R}\rangle, \tag{10}$$

$$\mathbf{E}_{mv}=\left|\mathbf{RVB}_{2,-\beta}\right\rangle+\left|\mathbf{RVB}_{0,\beta}\right\rangle. \tag{11}$$

Here, we take $\sigma=0$ in Supplementary Equation (6). According to Supplementary Equations (9-11), different polarization modes have different modulation factors or different orders, which demonstrates that the modulation symmetry of the light beam is broken. The breaking of the modulation symmetry provides the possibility of disentangling the inherent polarization modes in free space.

## Supplementary Note 3: Principle of polarization mode extraction

In this note, we present the entire theoretical principle of polarization mode extraction. As shown in Supplementary Figure 1, the experimental setup in Figure 2 can be simplified into a focusing system in which an $m$-order VVB is modulated by a phase-only spatial light modulator (SLM). The numerical aperture (NA) of the OL is 0.01.

Based on the Debye vectorial diffractive theory, the electric fields of the inherent polarization modes extracted in the focal region can be expressed as [1]

$$\mathbf{E} = -\frac{A}{\pi}\int_0^{2\pi}\int_0^{\alpha}\sin\theta\cos^{1/2}\theta T l_0(\theta)\mathbf{V}\exp(-ik\mathbf{s}\bullet\boldsymbol{\rho})\,d\theta d\varphi, \tag{12}$$

where $\theta$ and $\varphi$ are the convergent angle and azimuthal angle, respectively, and $A$ is a normalized constant. $\alpha = \arcsin(\text{NA}/\upsilon)$, where NA is the numerical aperture of the OL, and $\upsilon$ is the refractive index in the focusing space. The wavenumber is $k = 2\upsilon\pi/\lambda$, where $\lambda$ is the wavelength of the incident beam, and $\boldsymbol{\rho} = (r\cos\phi, r\sin\phi, z)$ denotes the position vector of an arbitrary field point. The unit vector along a ray is expressed as $\mathbf{s} = (-\sin\theta\cos\varphi, -\sin\theta\sin\varphi, \cos\theta)$. $T = \exp(i\psi)$ is the transmittance of the pupil filter P, where $\psi$ denotes the phase of the $m$-order VVB in Equation (7). $l_0(\theta)$ is the electric field amplitude of the incident beam, which can be expressed as [2]

$$l_0(\theta) = J_1(2\beta_0\frac{\sin\theta}{\sin\alpha})\exp[-(\beta_0\frac{\sin\theta}{\sin\alpha})^2], \tag{13}$$

where $\beta_0$ is the ratio of the pupil radius to the incident beam waist. $J_1(\bullet)$ is the Bessel function of the first kind with order 1.

According to Equation (2), the incident $m$-order VVB can be considered as the combination of two orthogonal polarization modes $|\mathbf{RVVB}_\sigma\rangle$ and $|\mathbf{LVVB}_\sigma\rangle$. The two polarization modes can be simplified into x and y linear polarization modes with $\sigma = 0$, respectively. That is, the electric field of the $m$-order VVB can be written as [3, 4]

$$\mathbf{E}_{\text{VVB}} = \cos(m\varphi)|\mathbf{x}\rangle + \sin(m\varphi)|\mathbf{y}\rangle, \tag{14}$$

where $|\mathbf{x}\rangle$ and $|\mathbf{y}\rangle$ denote the x and y linear polarization modes, respectively. Therefore, in Supplementary Equation (12), the propagation unit vector of the incident beam immediately after having passed through the lens is $\mathbf{V} = \cos m\varphi \mathbf{V}_x + \sin m\varphi \mathbf{V}_y$, where $\mathbf{V}_x$ and $\mathbf{V}_y$ are the electric vectors of $|\mathbf{x}\rangle$ and $|\mathbf{y}\rangle$, respectively, and can be written as [1]

$$\mathbf{V}_x = \begin{bmatrix} \cos\theta + (1-\cos\theta)\sin^2\varphi \\ -(1-\cos\theta)\sin\varphi\cos\varphi \\ \sin\theta\cos\varphi \end{bmatrix}; \mathbf{V}_y = \begin{bmatrix} -(1-\cos\theta)\sin\varphi\cos\varphi \\ 1-(1-\cos\theta)\sin^2\varphi \\ \sin\theta\sin\varphi \end{bmatrix}. \tag{15}$$

Eventually, the focal light intensity of the polarization modes extracted from the $m$-order VVB can be obtained using $I = |\mathbf{E}|^2$.

## Theoretical result

In the following simulations, NA=0.01, $\upsilon = 1$, and $\beta_0 = 1$. The unit of length in all figures is the wavelength $\lambda$, and the light intensity is normalized to the unit value.

### Difference between accurate and approximate wavefronts

According to Equation (4), polarization modes $\left|\mathbf{RVVB}_{2n+\sigma,-\beta}\right\rangle$ and $\left|\mathbf{RVVB}_{\sigma,\beta}\right\rangle$ are obtained by modulating the $m$-order VVB with $T = \cos(n\varphi - \beta)$, which is related to not only the phase of the VVB but also the amplitude. Thus, we rewrite this wavefront as $T = Amp \bullet \exp(i\phi)$, where $Amp = \text{Amplitude}(\cos(n\varphi - \beta))$ and $\phi = \text{Phase}(\cos(n\varphi - \beta))$. Supplementary Figure 2 shows the polarization modes created via the accurate wavefront of $T = \cos(n\varphi - \beta)$ in Supplementary Figure 2 (a) and the approximate wavefront of $T = \exp(i\phi)$ in Supplementary Figure 2 (e). Here, $n = 29$, and $\beta = 0$. Polarization modes $\left|\mathbf{RVVB}_{\sigma,\beta=0}\right\rangle$ with the order of $\sigma = 1$ are therefore extracted from the $m$=30-order VVB in the focal region; see Supplementary Figure 2 (b, f). Based on a comparison of the light intensities in Supplementary Figure 2 (c, d) with those in Supplementary Figure 2 (g, h), the polarization modes created based on Supplementary Figure 2 (a, e) are almost the same, with a deviation of only 0.26%. That is, the amplitude of $T = \cos(n\varphi - \beta)$ has little impact on the polarization mode extraction in the focal region. For this reason, we only take the phase of $T = \cos(n\varphi - \beta)$ into account in the paper by neglecting the amplitude of the $m$-order VVB.

### Polarization mode competition

Supplementary Figure 3 presents the breaking of the modulation symmetry by modulating the $m$-order VVB using $T = \exp(i\phi)$ with $\phi = \text{Phase}(\cos(n\varphi))$. Take Supplementary Figure 3 (h) as an example. The $m$=30-order VVB modulated by the phase $T = \exp(i\phi)$ with $n = 21$ shown in Supplementary Figure 3 (d) is divided into two polarization modes $\left|\mathbf{RVVB}_{\sigma,\beta}\right\rangle$ with the order $\sigma = 9,51$ and $\beta = 0$ in the focal region. The two modes not only possess identical total light intensities but also propagate coaxially in free space. As the propagation distance increases, the polarization modes with different orders exhibit different degrees of energy divergence, thereby leading to different positions in the focal region. Generally, a polarization mode of a lower order is located at the inner ring, while that of a larger order is located at the outer ring, as shown in Supplementary Figure 3 (h). This spatial separation implies the breaking of the modulation symmetry, which further disentangles the inherent polarization modes in free space.

As discussed above, the two inherent orthogonal polarization modes are two equal parts of the VVB. Although the total light intensities of both modes cannot be adjusted in free space, their energy densities are merely dependent on the order of the polarization modes. Generally, the lower the order of the polarization mode is, the higher the energy density. A higher energy density always leads to a larger maximum light intensity of the polarization mode. That is, the polarization mode at the inner ring is always brighter than that at the outer ring, as shown in Supplementary Figures 3 (e-h). Here, we define a parameter $\gamma = I_{m,outer} / I_{m,inner}$ to characterize the difference between the maximum light intensities of the inner ring $I_{m,inner}$ and outer ring $I_{m,outer}$. As shown in Supplementary Figure 3, different $\gamma$ can be obtained by different orders of polarization modes, which can further be adjusted at will by the parameter $\eta = \|m+n| - |-m+n\|$. Here, $|m+n|$ and $|-m+n|$ are the order of polarization modes at the outer and inner rings, respectively. If $\eta$ is sufficiently large, then a small $\gamma$ can always be obtained. With normalization, the polarization mode at the inner ring is retained by making the other mode negligible; see Supplementary Figures 3 (e-h). Here, we call this phenomenon polarization mode competition. Note that the circular polarization modes $|\mathbf{L}\rangle$ and $|\mathbf{R}\rangle$ in Equation (3) can also be analyzed in the same way.

**Theoretical result of polarization mode extraction**

Supplementary Figures 4 and 5 present the theoretical results of single and multiple polarization mode extraction in the focal region using an $m$=30-order VVB. Based on a comparison with Figures 3 and 5, the experimental results are consistent with the theoretical predictions in Supplementary Figures 4 and 5, respectively. In addition, we also verify the polarization conversion between $\left|\mathbf{RVVB}_{\sigma,\beta}\right\rangle$ and $\left|\mathbf{LVVB}_{\sigma,\beta}\right\rangle$ by adjusting the parameter $\beta$. Supplementary Figures 6 and 7 present the theoretical and experimental results of extracting multiple polarization modes $\left|\mathbf{RVVB}_{\sigma,\beta}\right\rangle$, where $\beta = -N\pi/8$ with

*N*=0, 1,…8 and (b) $\sigma = 0$, (g) $\sigma = 1$, (l) $\sigma = -1$, (q) $\sigma = 2$, and (v) $\sigma = -2$. When the light passes through the polarizer indicated by the purple arrows, $2\sigma$ petals with different $\beta$ are rotated; see Supplementary Figures 6 (h-j, m-o, r-t, w-y). In the case of $\sigma=0$ in Supplementary Figure 6 (b), $\left| \mathbf{RVVB}_{\sigma,\beta} \right\rangle$ is simplified into a linear polarization mode with the direction of $\beta$. For example, an x or y linear polarization mode can be obtained with $\beta = 0,\ \pi/2$. According to Malus' law, different polarization directions of a linear polarization mode manifest different light intensities. Thus, polarization modes with different brightnesses are obtained when the light passes through the polarizer indicated by the purple arrows in Supplementary Figures 6 (c-e). According to the experimental and theoretical results in Supplementary Figures 6 and 7, the polarization mode $\left| \mathbf{RVVB}_{\sigma,\beta} \right\rangle$ can be transformed into its orthogonal counterpart $\left| \mathbf{LVVB}_{\sigma,\beta} \right\rangle$ by merely adjusting the parameter $\beta$. Note that the phases for the extraction of the above polarization modes are shown in Supplementary Figures 6 (a, f, k, p, u), the parameters of which can be found in Supplementary Note 4.

# Supplementary Note 4: Parameters of all phases

## Figure 5 and Supplementary Figure 5

Phase of the $m$-order VVB:

$$\psi = \text{Phase}\left\{\sum_{j=1}^{N=9}[T_j \bullet \text{PF}(s_j, x_j, y_j, 0, \delta_j)]\right\},$$

where $l_1 = T_1 \bullet \text{PF}(s_1, 0, 0, 0, \delta_1)$; $l_2 = T_2 \bullet \text{PF}(s_2, d, 0, 0, \delta_2)$; $l_3 = T_3 \bullet \text{PF}(s_3, d, d, 0, \delta_3)$;

$l_4 = T_4 \bullet \text{PF}(s_4, 0, d, 0, \delta_4)$; $l_5 = T_5 \bullet \text{PF}(s_5, -d, d, 0, \delta_5)$; $l_6 = T_6 \bullet \text{PF}(s_6, -d, 0, 0, \delta_6)$;

$l_7 = T_7 \bullet \text{PF}(s_7, -d, -d, 0, \delta_7)$; $l_8 = T_8 \bullet \text{PF}(s_8, 0, -d, 0, \delta_8)$; $l_9 = T_9 \bullet \text{PF}(s_9, d, -d, 0, \delta_9)$; and $d = 800$.

## Parameters of the phase in Figure 5 (m) and Supplementary Figure 5 (m)

$\delta_j = 0$; $T_1 = \exp(-im\varphi)$; $T_j = \cos(m_j\varphi + \beta_j)$ with $\beta_j = 0$ and $m_j = m - j + 2$, $j = 2, 3 \ldots N$; $m = 30$;

$s_1 = 0.45$; $s_2 = 0.76$; $s_3 = 1.2$; $s_4 = 1.4$; $s_5 = s_6 = 1.6$; $s_7 = 1.8$; $s_8 = 2$; and $s_9 = 2.2$.

## Parameters of the phase in Figure 5 (n) and Supplementary Figure 5 (n)

$\delta_j = 0$; $T_1 = \exp(-im\varphi)$; $T_2 = \cos(m_2\varphi + \beta_2)$ with $m_2 = m$ and $\beta_2 = 0.38\pi$; $T_j = \cos(m_j\varphi + \beta_j)$ with

$\beta_j = 0$ and $m_j = m + j - 2$, $j = 3, 4 \ldots N$; $m = 30$;

$s_1 = s_2 = 0.5$; $s_3 = 1.2$; $s_4 = 1.4$; $s_5 = s_6 = 1.6$; $s_7 = 1.8$; $s_8 = 2$; and $s_9 = 2.2$.

## Parameters of the phase in Figure 5 (o) and Supplementary Figure 5 (o)

$\delta_j = 0$.

For $j = 1, 3, 5, 7, 9$:

$T_j = \cos(m_j\varphi + \beta_j)$ with $\beta_j = 0$ and $m_j = m + (j - 1)/2$.

For $j = 2, 4, 6, 8$:

$T_j = \cos(m_j\varphi + \beta_j)$ with $\beta_j = 0$ and $m_j = m - j/2$.

$s_1 = 0.55$; $s_2 = s_3 = 1$; $s_4 = s_5 = 1.25$; $s_6 = s_7 = 1.35$; and $s_8 = s_9 = 1.5$.

Final transmittance of the pupil filter P in Supplementary Equation (12):
$T = \exp(i\psi)$.

# Supplementary Figures 6 and 7

Phase of the $m$-order VVB:

$$\psi = \text{Phase}\left\{\sum_{j=1}^{N=8}[T_j \bullet \text{PF}(s_j, x_j, y_j, 0, \delta_j)]\right\},$$

where $x_j = d\cos\phi_j$ , $y_j = d\sin\phi_j$ and $d = 800$ ; $\phi_j = 2\pi(j-1)/N, j = 1,2,3...N$ ; $\delta_j = 0$ ; $s_j = 1$ ; and

$T_j = \cos(m_1\varphi + \beta_j)$ with $\beta_j = -\pi(j-1)/N, j = 1,2,3...N$ .

**Parameters of the phases in Supplementary Figures 6 (a) and 7 (a)**

$m_1 = m = 30$ .

**Parameters of the phases in Supplementary Figures 6 (f) and 7 (f)**

$m_1 = m - 1$ .

**Parameters of the phases in Supplementary Figures 6 (k) and 7 (k)**

$m_1 = m + 1$ .

**Parameters of the phases in Supplementary Figures 6 (p) and 7 (p)**

$m_1 = m - 2$ .

**Parameters of the phases in Supplementary Figures 6 (u) and 7 (u)**

$m_1 = m + 2$ .

Final transmittance of the pupil filter P in Supplementary Equation (12):
$T = \exp(i\psi)$ .

**Supplementary References**